# Reduced conditional dynamic of quantum system under indirect quantum measurement

[1]Trifanov A.I., [2]Miroshnichenko G.P.,

Saint Petersburg National Research University of Information Technologies, Mechanics and Optics, 197101, Saint Petersburg, Kronverksky avenue, 49

e-mail: [1]alextrifanov@gmail.com, [2]gpmirosh@gmail.com

**Abstract:** In this report, we study the reduced conditional dynamics of a quantum system in the case of indirect quantum measurement. The detector's microscopic part (pointer) interacts with the measured system (target) and the environment, which results in a nonunitary interaction between target and pointer. The quantum state evolution conditioned by the measurement result is under investigation. Particularly, we are interested in explicit analytical expressions for the conditional evolution superoperators and basic information characteristics of this measurement process, which is applied to the cavity mode photodetection problem.
**PAC's:** 42.50.Ar, 42.50.Lc, 42.50.Pq

## 1. INTRODUCTION

An interaction between a quantum system and the environment is one of the fundamental problems in Quantum Physics. In light of quantum communication and computation development [1, 2], the evolution of monitored quantum systems becomes very important, though the guidelines are difficult at present. Particularly, quantum measurement [3], decoherentization and superselection [4] problems, which are significant, are closely related to the theory of open quantum system dynamics [5].

There are several approaches, which deal with quantum system evolution in the presence of the environment, e.g. the optical master equation [6], the model of quantum Brownian motion [7] and their generalizations for the cases of non Markovian dynamics [8]. In this context, different measurement models were proposed: indirect measurements, weak measurements, nondemolition measurements, etc. [3]. From a formal point of view, one can assign at least three formulations, which reflect the detector's role in the quantum measurement process [9, 10]: the model of projection measurement based on the von Neumann postulate [11]; the language of "effects" and "operations" [12] and formalism based on the Feynman path integral [13, 14].

The technique of direct projective measurement is the first proposed and the simplest way to estimate the state of a quantum system. Unfortunately, during this process, we often have to cancel the system of interest. The idea of indirect measurement is to organize an interaction between the measured system and the detector's microscopic part (pointer) followed by pointer state detection. This gives a wide opportunity for quantum state preparation, unsharp and weak measurements, realization of quantum logical operations [15, 16], quantum state [17] and quantum process [18, 19] tomography.

Formally, the indirect measurement process may be described by parameterization $\left(\rho_P, U_\tau, \{\Pi_x\}\right)$ of the initial projective measurement. Here, $\rho_P$ is initial pointer state, $U_\tau$ is unitary operator and $\{\Pi_x\}$ - set of projector valued measures. This may be completed by the canonical Naimark extension [20]. This technique was successfully applied in [21] to describe the Stern-Gerlach experiment with unsharp measurement. Another way is to start from an arbitrarily suitable pointer and comparatively simple interaction and construct Positive Operator Valued Measures (POVM) corresponding to this kind of generalized detection. This method was used for the description of indirect measurement on trapped ions [22], obtaining quantum control and quantum gate realization in the QED cavity [16].

In fact, the efficiency of a measurement device is usually not exactly equal to identity [23] (not to speak of the photodetection and particularly IR region, where this quantity is quite different than unity). One can define at least two factors, which are participating in this process. The first one is a classical and quantum stochastic process, which governs the behavior of the measurement apparatus. This factor may be taken into account by introducing the corresponding phenomenological probability distributions for detection events and errors [24].

The second one is a nonunitary interaction between the measured system and the microscopic part of detector (pointer), which may take place due to the interaction with the environment. This process may change detection statistics significantly. To consider it, one should construct the parameterization scheme using nonunitary evolution. Namely, instead of triple $(\rho_P, U_\tau, \{\Pi_x\})$, one should use $(\rho_P, \mathcal{U}_\tau, \{\Pi_x\})$, where $\mathcal{U}_\tau = \exp(-i\mathcal{L}\tau)$ and $\mathcal{L}$ is the evolution generator.

Here, we examine the situation of nonunitary evolution considering the intracavity interaction between two level atom and cavity quantum mode during the process of indirect electromagnetic field measurement. We model the process, when two level atom passes through the cavity and "collects" information about the state of quantum mode. Just after the interaction, the atomic state is determined in a selective detector, which gives one of two possible alternatives: atom is found in its ground state $|g\rangle_A$ or in an exited state $|e\rangle_A$. From these results, the state of cavity quantum mode may be calculated.

The structure of this paper is organized as follows: *in section 2* we give a brief review about the conception of quantum measurement process. *Section 3* consists of the description for the model problem of intracavity quantum mode photodetection and presents the master equation in superoperator form. *In section 4* the system of differential equations for conditional superoperators is obtained and solved in two extreme cases: strong and weak relaxation limits. The basic information characteristics are presented *in section 5*. Here we show them us a function of interaction time and discuss the basic results. *Section 6* concludes the paper.

## 2. PARAMETRIZATION OF THE QUANTUM MEASUREMENT PROCESS

The goal of this section is to briefly give a review about the conception of the quantum measurement process. We will start from the ordinary case of projection measurement and define the set of projection valued measures. Then, the concept of indirect measurement will be presented and the corresponding generalization of detectors measures (POVM) will be described. At the end of this section we will introduce a parametrization scheme for indirect measurement followed by interaction between the pointer and the environment.

## 2.1 Von Neumann measurement

For complete specification of the measurement apparatus, a full set of projector valued measures $\{\Pi_x\}$ should be known. Let $B_S$ be the outcome space of this device, then for every detection result $r \in B_S$ the probability distribution:

$$p_r = Tr_S(\rho_S \cdot \Pi_r), \tag{1}$$

may be obtained. Here, $\rho_S$ is the quantum state of a measured system and $Tr_S$ is the trace over its state space $H_S$. The completeness condition gives us:

$$\sum_r \Pi_r = I_S, \tag{2}$$

where $I_S$ is the operator identity.

Following the von Neumann postulate, we can determine the state of the system just after the detection process with certain outcome $r$:

$$\rho_S \to \rho_S^r = \frac{\Pi_r \rho_S \Pi_r}{Tr[\rho_S \cdot \Pi_r]} = \frac{\Lambda_r \rho_S}{Tr[\Lambda_r \rho_S]}. \tag{3}$$

Here, $\Lambda_r$ is state transformer. It should be emphasized that there is an ambiguity in the definition of transformers $\Lambda_r$, which reflects the fact that different detectors may have identical measurement statistics. So, we have to mention the equivalence class of measurement devices.

## 2.2 Indirect measurement. Unitary evolution

Now, we can assume that the projection measurement described above is used for state detection of some auxiliary system (pointer), which had interacted with our system just before the measurement. Let us denote the state space of ancilla by $H_A$ and the outcome space of ancilla state detector by $B_A$. The triplet of parameters $(\rho_P, U_\tau, \{\Pi_x\}_A)$ describes the change of the system state in the presence of detector which outcome is ignored:

$$\rho_S \to \tilde{\rho}_S = Tr_A\left[U_\tau(\rho_S \otimes \rho_P)U_\tau^\dagger\right]. \tag{4}$$

Here, $Tr_A$ is a trace operation in $H_A$.
Additional information, obtained by using the measurement result $r \in B_A$ of pointer state, allows one to determine the conditional state of monitored system. Let $\Pi_r = |\varphi_r\rangle\langle\varphi_r|$, $|\varphi_r\rangle \in H_A$, then the following mapping describes the conditional evolution of quantum system:

$$\rho_S \to \rho_S^r = \langle\varphi_r|\left[U_\tau(\rho_S \otimes \rho_P)U_\tau^\dagger\right]|\varphi_r\rangle = \Xi_r \rho_S, \tag{5}$$

Due to the first Kraus representation theorem, the action of transformer $\Xi_r$ on arbitrary state $\rho_S$ may be written as follows:

$$\Xi_r \rho_S = \sum_j M_j^r \rho_S M_j^{r\dagger}, \qquad (6)$$

where $M_j^r$ is bounded (Kraus) operators. If $\rho_P = |in\rangle\langle in|$, we can write for them:

$$M_j^r = a(j|r) \langle \varphi_j | U_\tau | in \rangle, \qquad (7)$$

where $a(j|r)$ is the amplitude of conditional probability to find pointer in state $\varphi_j$ if detection result $\varphi_r$ is obtained (for an imperfect measurement apparatus).

There is a direct but nontrivial way to calculate the Kraus operators in the case of unitary evolution between system and pointer. Let $H_{int}$ be their interaction Hamiltonian. Substituting the evolution operator decomposition:

$$U(t) = \sum_{i,r \in B_A} |\varphi_i\rangle\langle\varphi_r| M_{i,r}(t), \qquad (8)$$

into the Schrödinger equation:

$$i\frac{d}{dt}U(t) = H_{int}U(t), \; U(0) = I, \qquad (9)$$

we can obtain the following system of differential equations, which determine the time evolution of Kraus operators $M_{i,r}(t)$ (index $i$ corresponds to the initial atomic state whereas $r$ corresponds to detected state):

$$i\frac{d}{dt}M_{i,r}(t) = \sum_{k \in B_A} \langle \varphi_i | H_{int} | \varphi_k \rangle \cdot M_{k,r}(t).$$

Its complexity depends on the algebra structure which obey the operators $\langle \varphi_i | H_{int} | \varphi_r \rangle$. For some cases, we can find exact analytical solution for them, but in general, it is a quite a nontrivial problem.

**2.3 Indirect measurement. Nonunitary evolution**

Interaction between the pointer and the environment leads to a nonunitary common reduced evolution of both target and pointer. To take it into account, the triplet $(\rho_P, \mathcal{U}_\tau, \{\Pi_x\})$ is introduced. For convenience, in future discussion, the superoperator representation for density matrix is used:

$$\rho_P = |in\rangle\langle in| = |in\rangle\rangle. \qquad (10)$$

In the following, we will assume that a secular approximation is established. Namely, the typical time scale of the intrinsic evolution of the pointer is large compared to the time over which it's state varies appreciably due to interaction with the environment. Under this approximation, description of the measurement process formally is the same as in the previous subsection. Namely, secular approximation allows the following decomposition of superoperator $\mathcal{U}_\tau$:

$$\mathcal{U}_\tau = \sum_{i,r \in B} |\varphi_i\rangle\rangle\langle\langle\varphi_r| \mathfrak{M}_{i,r}(\tau), \qquad (11)$$

where we write $\mathfrak{M}_{i,r}$ instead of $M_{i,r}$ underlining that $\mathfrak{M}_{i,r}$ is superoperator (transformer) and $\mathfrak{M}_{i,r} = \Xi_r$ for initial conditions, indexed by $i \in B_A$. Substitution of equation (11) into the master equation for the diagonal density matrix (rotating wave approximation):

$$i\dot{\rho}^d(t) = \mathcal{L}\rho^d(t)$$

gives the following system of differential equations for superoperators $\mathfrak{M}_{s,r}$:

$$i\dot{\mathfrak{M}}_{i,r}(t) = \sum_{k \in B_A} \langle\langle \varphi_i | \mathcal{L} | \varphi_k \rangle\rangle \mathfrak{M}_{k,r}(t). \tag{12}$$

For every fixed $i$ we obtain the closed system for $\dot{\mathfrak{M}}_{i,r}$. In the next section, this formalism will be applied to the indirect photodetection problem.

## 3. MASTER EQUATION

Here, we will apply the indirect measurement model, followed by nonunitary interaction, to the problem of intracavity mode photodetection. The master equation in this case has the Lindblad form:

$$i\dot{\rho} = \mathcal{L}\rho = [H_{int}, \rho] + i\mathcal{D}_A\rho, \tag{13}$$

where $H_{int}$ is the Jaynes-Cummings Hamiltonian:

$$H_{int} = \Omega\sigma_+ a \exp(i\Delta t) + \Omega^*\sigma_- a^\dagger \exp(-i\Delta t), \tag{14}$$

$\mathcal{D}_A$ is the atomic "dissipater"

$$\mathcal{D}_A\rho = \gamma_{ge}/2(2\sigma_-\rho\sigma_+ - \sigma_+\sigma_-\rho - \rho\sigma_+\sigma_-) + \gamma_{eg}/2(c.c.), \tag{15}$$

and the following notations are used: $\Omega$ is a coupling between atom and mode, $\sigma_+ = |e\rangle\langle g|$ and $\sigma_- = |g\rangle\langle e|$ are atomic operators, $a$ is an annihilation operator of cavity mode, $\Delta$ is detuning, $\gamma_{ge}$ and $\gamma_{eg}$ are atomic population relaxation rates. Also it is more convenient to write the dissipater in superoperator form:

$$(\mathcal{D}_A\rho)_{j,k} = \begin{cases} -\rho_{jk}\Gamma_{jk}, \Gamma_{jk} = \Gamma_{kj} = \Gamma, j \neq k; \\ \sum_{q \in \{g,e\}} (\gamma_{qk}\rho_{qq} - \gamma_{kq}\rho_{kk}), j = k, \end{cases} \tag{16}$$

where we introduce nonzero phase relaxation rate $\Gamma$.

Let $B_A = \{g, e\}$ be a set of results of atomic state detection. Here, we will assume the ideal (projective) measurement of pointer state. Using density operator decomposition:

$$\rho_{AF}(t) = \sum_{\mu,\nu \in B_A} |\mu\rangle\langle\nu| \otimes \rho_{\mu\nu}(t), \tag{17}$$

in (13), one can obtain a system of differential equations for matrix elements of the density operator $\rho_{AF}$ in atomic basis:

$$\dot{\rho}_{gg} = -i\left(\Omega^* a^\dagger \rho_{eg} e^{-i\Delta t} - \Omega \rho_{ge} a e^{i\Delta t}\right) + \gamma_{eg}\rho_{ee} - \gamma_{ge}\rho_{gg},$$
$$\dot{\rho}_{ge} = -i\left(\Omega^* a^\dagger \rho_{ee} e^{-i\Delta t} - \Omega^* \rho_{gg} a^\dagger e^{-i\Delta t}\right) - \Gamma_{ge}\rho_{ge},$$
$$\dot{\rho}_{eg} = -i\left(\Omega a \rho_{gg} e^{i\Delta t} - \Omega \rho_{ee} a e^{i\Delta t}\right) - \Gamma_{eg}\rho_{eg}, \quad (18)$$
$$\dot{\rho}_{ee} = -i\left(\Omega a \rho_{ge} e^{i\Delta t} - \Omega^* \rho_{eg} a^\dagger e^{-i\Delta t}\right) + \gamma_{ge}\rho_{gg} - \gamma_{eg}\rho_{ee}.$$

The transformation $\rho_{ge} = \tilde{\rho}_{ge} e^{-i\Delta t}$, $\rho_{eg} = \tilde{\rho}_{eg} e^{i\Delta t}$ (in the following, the tilde will be dropped) and secular approximation (rotating wave approximation) $\Gamma \gg \Omega$ $\left(\dot{\rho}_{ge} = \dot{\rho}_{eg} = 0\right)$, gives the following closed system of differential equations for diagonal elements $\rho_{gg}(t)$ and $\rho_{ee}(t)$ of density matrix:

$$\dot{\rho}_{gg} = \kappa\Gamma\left(2a^\dagger \rho_{ee} a - \rho_{gg} a^\dagger a - a^\dagger a \rho_{gg}\right) - i\kappa\Delta\left[a^\dagger a, \rho_{gg}\right] + \gamma_{eg}\rho_{ee} - \gamma_{ge}\rho_{gg},$$
$$\dot{\rho}_{ee} = \kappa\Gamma\left(2a \rho_{ee} a^\dagger - \rho_{gg} a a^\dagger - a a^\dagger \rho_{gg}\right) - i\kappa\Delta\left[a^\dagger a, \rho_{ee}\right] + \gamma_{ge}\rho_{gg} - \gamma_{eg}\rho_{ee}. \quad (19)$$

which may be written as:
$$\dot{\rho}^d(t) = \mathcal{L}'\rho^d(t), \quad (20)$$

where $\rho^d(t) = \text{diag}\{\rho_{gg}(t), \rho_{ee}(t)\}$ and $\kappa = |\Omega|^2/(\Gamma^2 + \Delta^2)$.

To obtain the analytical expression for $\mathcal{L}'$ in superoperator form, let us introduce the following superoperators, which act on the density matrix of cavity field:

$$K_0 \rho_F = \frac{1}{2}\left(a^\dagger a \rho_F + \rho_F a a^\dagger\right),\ K_+ \rho_F = a^\dagger \rho_F a,\ K_- \rho_F = a \rho_F a^\dagger,\ N\rho_F = \left[a^\dagger a, \rho_F\right], \quad (21)$$

and obey the commutation relations of $SU(1,1)$ algebra:

$$\left[K_-, K_+\right] = 2K_0,\ \left[K_0, K_+\right] = K_+,\ \left[K_0, K_-\right] = -K_-,$$
$$\left[K_0, N\right] = \left[K_+, N\right] = \left[K_-, N\right] = 0. \quad (22)$$

Also, for the following, it is convenient to introduce atomic superoperators $\theta_{jk}$:

$$\theta_{jk}\rho = |j\rangle\rangle\langle\langle k|\rho = |j\rangle\langle k|\rho|k\rangle\langle j|. \quad (23)$$

With these notations, we can write evolution equation in Lowville form (20) with

$$\mathcal{L}' = \mathcal{L}_0 + \mathcal{L}_r, \quad (24)$$

where

$$\mathcal{L}_0 = -\kappa\Gamma\left[K_0\left(\theta_{gg} + \theta_{ee}\right) + \frac{1}{2}\left(\theta_{ee} - \theta_{gg}\right) - K_+\theta_{ge} - K_-\theta_{eg}\right] - i\kappa\Delta N\left(\theta_{gg} - \theta_{ee}\right),$$
$$\mathcal{L}_r = \sum_{j,k}\gamma_{jk}\left(\theta_{kj} - \theta_{jj}\right). \quad (25)$$

## 4. SUPEROPERATORS OF CONDITIONAL EVOLUTION

To solve Eq. (20) with the evolution generator taken from (25), we will use evolution superoperator decomposition (11) in the following form:

$$\rho(t) = \mathfrak{A}(t)\rho(0) = \sum_{i,j \in B_A} \mathfrak{M}_{ij}(t) \otimes |\varphi_i\rangle\rangle\langle\langle\varphi_j|\rho(0). \tag{26}$$

Substituting the first part of this equality into (20) gives the differential equation for superoperator $\mathfrak{A}(t)$:

$$\dot{\mathfrak{A}}(t) = (\mathcal{L}_0 + \mathcal{L}_r)\mathfrak{A}(t), \quad \mathfrak{A}(0) = \mathfrak{I}. \tag{27}$$

Here, $\mathfrak{I}$ is the identical superoperator. Then, using the second part of equality (26) in (27), we can write two separate systems of differential equations. One of these determines superoperators $\mathfrak{M}_{gg}$ and $\mathfrak{M}_{eg}$, which describe the conditional evolution when atom was prepared in its ground state $|g\rangle$ and detected in states $|g\rangle$ and $|e\rangle$ correspondingly. Other superoperators $\mathfrak{M}_{ge}$ and $\mathfrak{M}_{ee}$ describe the measurement process with an atom prepared in the excited state $|e\rangle$. We will solve only the first of these systems, because the method for solving the second one is identical. For the first one, we can write (dependence of $t$ is dropped):

$$\begin{cases} \dot{\mathfrak{M}}_{gg} = -(\alpha K_0 + \beta + \gamma_{ge})\mathfrak{M}_{gg} + (\alpha K_+ + \gamma_{eg})\mathfrak{M}_{eg}, \\ \dot{\mathfrak{M}}_{eg} = (\alpha K_- + \gamma_{ge})\mathfrak{M}_{gg} - (\alpha K_0 - \beta + \gamma_{eg})\mathfrak{M}_{eg}. \end{cases} \tag{28}$$

Here, $\alpha = \kappa\Gamma$ and $\beta = \kappa(i\Delta N - \Gamma/2)$. This system may also be represented in a more symmetrical way with substitution:

$$\begin{aligned} \mathfrak{M}_{gg} &= \exp\left[-(\alpha K_0 + \beta + \gamma_{ge})\right]\mathfrak{M}'_{gg}, \\ \mathfrak{M}_{eg} &= \exp\left[-(\alpha K_0 - \beta + \gamma_{eg})\right]\mathfrak{M}'_{eg}, \end{aligned} \tag{29}$$

in use. In this case it may be written as follows:

$$\begin{cases} \dot{\mathfrak{M}}'_{gg} = (e^{\alpha t}K_+ + \gamma_{eg})e^{(2\beta + \gamma_{ge} - \gamma_{eg})t}\mathfrak{M}'_{eg}, \\ \dot{\mathfrak{M}}'_{eg} = (e^{-\alpha t}K_- + \gamma_{ge})e^{-(2\beta + \gamma_{ge} - \gamma_{eg})t}\mathfrak{M}'_{gg}. \end{cases} \tag{30}$$

It is a nontrivial problem to get the exact analytical solutions for systems (28) or (30). Here, we use two extreme cases of strong and weak relaxation to obtain approximate expressions for the superoperators $\mathfrak{M}_{gg}$ and $\mathfrak{M}_{eg}$ from perturbation theory up to the first order:

$$\begin{aligned} \mathfrak{M}_{gg} &= \mathfrak{M}_{gg}^{(0)} + \mathfrak{M}_{gg}^{(1)}, \\ \mathfrak{M}_{eg} &= \mathfrak{M}_{eg}^{(0)} + \mathfrak{M}_{eg}^{(1)}, \end{aligned} \tag{31}$$

leaving the exact analytical solutions for subsequent research.

**4.1 Perturbation theory. Strong relaxation**

In the secular approximation, which we used to obtain (28), we assumed that $|\Omega| \ll \Gamma$, which leads to $\kappa \ll 1$. Let us assume now that $\gamma_{ge} < \kappa\Gamma \approx \kappa\Delta \ll \gamma_{eg}$, which means that the low temperature approximation is used. It gives the following system for the zero order approximation:

$$\begin{cases} \dot{\mathfrak{M}}_{gg}^{(0)} = \gamma_{eg}\mathfrak{M}_{eg}^{(0)}, \\ \dot{\mathfrak{M}}_{eg}^{(0)} = -\gamma_{eg}\mathfrak{M}_{eg}^{(0)}. \end{cases} \qquad (32)$$

It has a simple analytical solution:

$$\mathfrak{M}_{gg}^{(0)} = \mathfrak{I}, \ \mathfrak{M}_{eg}^{(0)} = 0, \qquad (33)$$

for the first order, the system has the form:

$$\begin{cases} \dot{\mathfrak{M}}_{gg}^{(1)} = \gamma_{eg}\mathfrak{M}_{eg}^{(1)} - (\alpha K_0 + \beta + \gamma_{ge}), \\ \dot{\mathfrak{M}}_{eg}^{(1)} = -\gamma_{eg}\mathfrak{M}_{eg}^{(1)} + (\alpha K_- + \gamma_{ge}), \end{cases} \qquad (34)$$

and solution:

$$\mathfrak{M}_{gg}^{(1)}(t) = (\alpha K_- + \gamma_{ge})(e^{-\gamma_{eg}t} - 1)/\gamma_{eg} - [\alpha(K_0 - K_-) + \beta]t,$$
$$\mathfrak{M}_{eg}^{(1)}(t) = (\alpha K_- + \gamma_{ge})(1 - e^{-\gamma_{eg}t})/\gamma_{eg}. \qquad (35)$$

**4.2 Perturbation theory. Weak relaxation**

Here we examine another extreme case, assuming that $\gamma_{ge} < \gamma_{eg} \ll \kappa\Gamma \approx \kappa\Delta$, so that in the zero order of perturbation theory, we get the following system:

$$\begin{cases} \dot{\mathfrak{M}}_{gg}^{(0)} = -(\alpha K_0 + \beta)\mathfrak{M}_{gg}^{(0)} + \alpha K_+\mathfrak{M}_{eg}^{(0)}, \\ \dot{\mathfrak{M}}_{eg}^{(0)} = \alpha K_-\mathfrak{M}_{gg}^{(0)} - (\alpha K_0 - \beta)\mathfrak{M}_{eg}^{(0)}. \end{cases} \qquad (36)$$

From straightforward calculations, one can obtain:

$$\ddot{\mathfrak{M}}_{eg}^{(0)} + \alpha(2K_0 + 1)\dot{\mathfrak{M}}_{eg}^{(0)} + (\alpha^2 C - \beta(\alpha + \beta))\mathfrak{M}_{eg}^{(0)} = 0, \qquad (37)$$

where $C = K_0^2 - K_0 - K_+K_-$ is the Casimir operator. Notice, that all coefficients in (37) are commute, so we can solve it as an ordinary second order differential equation. The characteristic roots are:

$$\mu_{1,2} = \frac{1}{2}\left(\pm\sqrt{D} - \alpha(2K_0 + 1)\right), \qquad (38)$$

where $D = 2\alpha^2(K_+K_- + K_-K_+) + (\alpha + 2\beta)^2 + 4\alpha^2 K_0$, and the solution may be written as:

$$\mathfrak{M}_{eg}^{(0)}(t) = \alpha\left(e^{\mu_1 t} - e^{\mu_2 t}\right)\left(\sqrt{D}\right)^{-1}K_-. \qquad (39)$$

To obtain expression for $\mathfrak{M}_{gg}(t)$, we use the following substitution:

$$\dot{\mathfrak{M}}_{gg}'^{(0)} = \alpha K_+ e^{(\alpha-a)t}\mathfrak{M}_{eg}^{(0)},$$

which leads to

$$\mathfrak{M}_{gg}^{(0)} = e^{-(\alpha K_0 + \beta)t}\left\{1 + \alpha^2 K_+\left[\lambda_1^{-1}(e^{\lambda_1 t} - 1) + \lambda_2^{-1}(e^{\lambda_2 t} - 1)\right]\sqrt{D^{-1}}K_-\right\},$$

where $\lambda_{1,2} = \mu_{1,2} + \alpha - a = \dfrac{\alpha \pm \sqrt{D}}{2} + \beta$.

In the first order of perturbation theory, we get the following system:

$$\begin{cases} \dot{\mathfrak{M}}^{(1)}_{gg} = -(\alpha K_0 + \beta)\mathfrak{M}^{(1)}_{gg} + \alpha K_+ \mathfrak{M}^{(1)}_{eg} + \gamma_{eg}\mathfrak{M}^{(0)}_{eg}, \\ \dot{\mathfrak{M}}^{(1)}_{eg} = \alpha K_- \mathfrak{M}^{(1)}_{gg} - (\alpha K_0 - \beta)\mathfrak{M}^{(1)}_{eg} - \gamma_{eg}\mathfrak{M}^{(0)}_{eg}. \end{cases} \quad (40)$$

Repeating the same procedure as in the zero order solution and using method of variation of parameters, one can obtain the following first order solution:

$$\mathfrak{M}^{(1)}_{eg}(t) = \gamma_{eg} D^{-1}\left\{(\lambda_1 + \lambda_2)\mathfrak{M}^{(0)}_{eg}(t) - \alpha t\left(\lambda_1 e^{\mu_1 t} - \lambda_2 e^{\mu_2 t}\right)K_-\right\} \\ + \gamma_{eg}\int_0^t \mathfrak{M}^{(0)}_{eg}(t-\tau)\mathfrak{M}^{(0)}_{eg}(\tau)d\tau, \quad (41)$$

$$\mathfrak{M}^{(1)}_{gg}(t) = \alpha^2 \gamma_{eg} K_+ D^{-3/2}\Big[(\lambda_1+\lambda_2)\left(e^{\mu_1 t}/\lambda_1 - e^{\mu_2 t}/\lambda_2\right) \\ + (\lambda_1+\lambda_2)(\mu_1-\mu_2)e^{-(\alpha K_0+\alpha+\beta)t}/\lambda_1\lambda_2 - (\mu_1-\mu_2)\left(e^{\mu_1 t}+e^{\mu_2 t}\right)t\Big]K_- \quad (42) \\ + \alpha\gamma_{eg} K_+ \int_0^t\int_0^\tau e^{-(\alpha K_0+\alpha+\beta)(t-\tau)}\mathfrak{M}^{(0)}_{eg}(\tau-\theta)\mathfrak{M}^{(0)}_{eg}(\theta)d\theta d\tau.$$

## 5. INFORMATION CHARACTERISTICS OF PHOTODETECTION PROCESS

Here, the basic information characteristics of the photodetection process will be investigated in the bounds of the model discussed above. The following quantities are of most interest: probability of certain detection result $r \in \{g,e\}$:

$$P_r(t) = Tr_F\left[\mathfrak{M}_{rg}(t)\rho_F\right], \quad (43)$$

information gain $I_r = -\Delta H$ as a measure of entropy change resulting from photodetection:

$$\Delta H = \rho_F^r \log \rho_F^r - \rho_F \log \rho_F \quad (44)$$

and fidelity:

$$F_r = \sqrt{\sqrt{\rho_F}\rho_F^r\sqrt{\rho_F}}, \quad (45)$$

which characterizes the state change caused by the measurement process.

These quantities are shown on Fig.1 - Fig.4 as functions of time interaction $\tau$ for different dimensions of cavity mode state space. Fig. 1 and Fig. 2 show the results obtained from the strong relaxation approximation, while Fig 3 and Fig 4 show these dependences for the case of weak relaxation limit. For each approximation, two initial states are tested: completely mixed state $\rho_F = 1/d$ and Fock state $\rho_F = |n\rangle\langle n|$. Here, $d$ is the dimension of the cavity mode state space and $n \leq d$ is the number of photons in cavity mode (cases $d = 2,4,6$ and $n = 1,3,5$ are depicted). We investigate the case of conditional reduced density matrix evolution, which corresponds to the detection result, $r = g$. The top rows of the graphs in each figure represent information characteristics in different approximations, while the bottom rows in Fig.1 and Fig.2 show the field density matrix elements in certain moment ($\Omega\tau = 0,10,50$) in the case $d = 4$ (or $n = 3$).

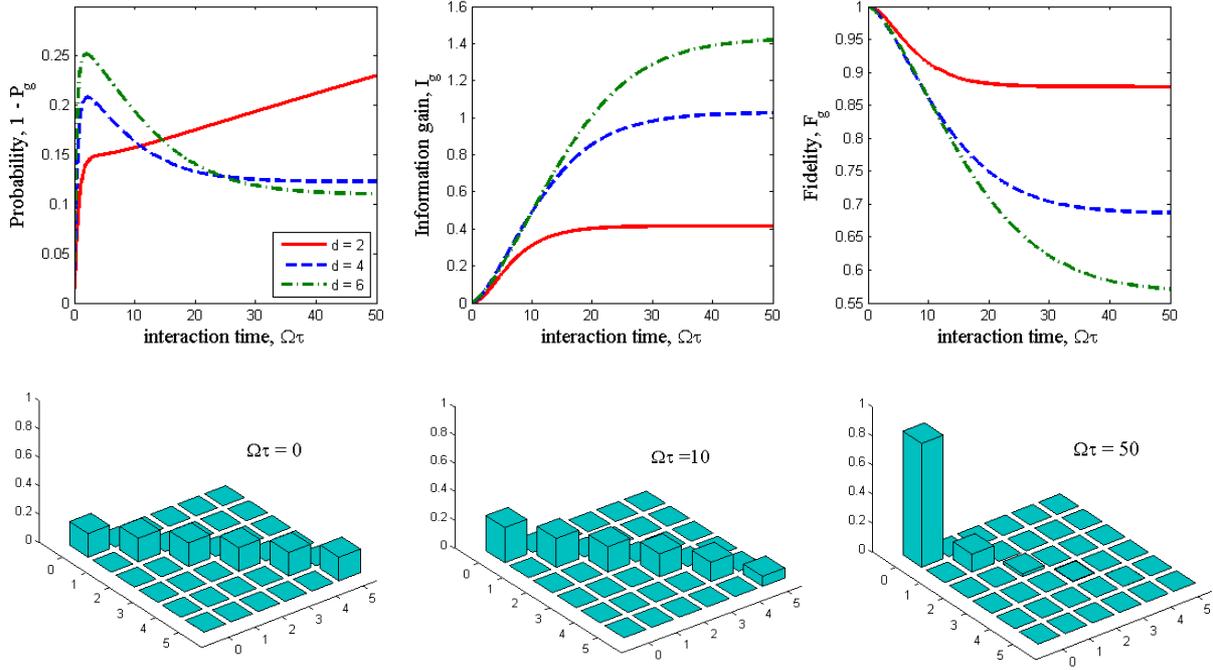

**Fig. 1** Ground state detection: probability, information gain and fidelity as a functions of interaction time for mixed state $\rho_F = 1/d$ (top); density matrix elements ($d = 4$) at the moments $\Omega\tau = 0, 10, 50$ (bottom). Strong relaxation approximation.

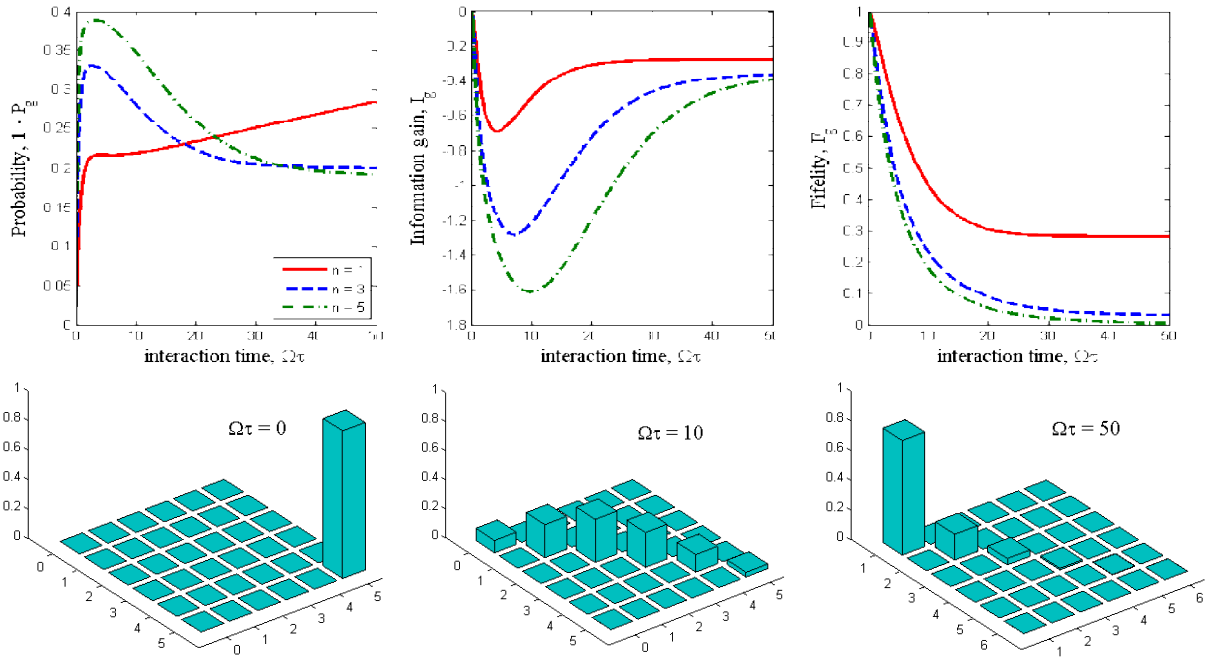

**Fig. 2** Ground state detection: probability, information gain and fidelity as a functions of interaction time: for Fock state $\rho_F = |n\rangle\langle n|$ (top); density matrix elements ($n = 3$) at the moments $\Omega\tau = 0, 10, 50$ (bottom). Strong relaxation approximation.

The first graph in each of the top rows shows the probability $1 - P_g = P_e$ ~~not~~ to detect atom in its non-ground state. In strong relaxation, the limit (Fig. 1 and Fig. 2) curves corresponding to interaction with several photons (blue dotted and green chain lines) tend to approach a constant nonzero value due to the nonzero atomic spontaneous excitation. In the case of interaction with only one photon in the cavity mode (red solid line) the probability to detect an atom in its excited state increases with time because of the comparatively weak atom-field interaction. For the weak relaxation limit, all curves tends to some constant value.

The second graph in each of top rows shows the information gain as a function of time. The behavior of this dependence for the cases of initially mixed (Fig. 1 and Fig. 3) and Fock states (Fig.2 and Fig.4) are completely different. For the initial mixed state in strong relaxation limit, all curves increase monotonically because this state has maximum entropy and tend to some constant value, which corresponds to the final vacuum state (with zero entropy). For the mode prepared in the Fock state and in the same limit, initial entropy is zero and rises due to common evolution with the atom. Finally, all curves tend to the vacuum value, which surely has zero entropy, but due to the nonzero spontaneous atomic excitation, the resulting state has a constant non-vanishing value of entropy. These dependences for the weak relaxation limit may be explained in the same way.

Finally, the third graph in each of the top rows shows the detection fidelity as a function of time. It is interesting to investigate the stationary limit of these dependences in strong relaxation approximation. In this approximation, all curves decay monotonically, which corresponds to an irreversible state change. For the field prepared in the completely mixed state, the stationary value is nonzero for all cases because the mixed state and the final vacuum state are nonorthogonal. For the case of Fock state initially prepared, the curves tends to approach zero with the increased ~~of~~ photon numbers. This may be explained by the orthogonality of the initial and final states. But, due to spontaneous excitation, the probability to detect an atom in its excited state is nonzero and superoperators of conditional evolution transform the density matrix to its final state, which is not orthogonal to its initial state. That is why all curves tend to approach small but nonzero value.

In the weak relaxation limit for the considered interaction times, stationary limits are not obtained. In this approximation (see bottom rows in Fig.3 and Fig.4), we compare the numerical simulation and the analytical solution (42) results for the case $d = 4$ ($n = 3$). Here, we used the quadrature method to calculate the integral in (42).

For numerical modeling, the following values of parameters are used (in unit $\gamma_{eg} = 1$): $\gamma_{ge} = 0.1$, $\gamma_{eg} = 1$, $\Gamma = 2$, $\Delta = 0.5$ and $\Omega = 0.7$ (for strong relaxation limit); $\gamma_{ge} = 0$, $\gamma_{eg} = 0.01$, $\Gamma = 2$, $\Delta = 0.5$ and $\Omega = 0.7$ (for weak relaxation limit).

## 6. CONCLUSION

The parametrization of the measurement process in the presence of a nonunitary evolution was investigated. Interaction between the pointer and the environment was discussed and applied to the quantum photodetection problem. Basic characteristics of measurement process as a function of interaction time for two approximations were obtained. Properties of superoperators corresponded to different measurement results may be used for finding special regimes of detection: measurement without state or entropy change from one side and detection with best information gain from another. The generalization of previous description beyond secular approximation (quantum Brownian motion) will be considered in the future. It is attractive to obtain detector superoperators from evolution generator, though it requires deep analysis of their algebraic properties.

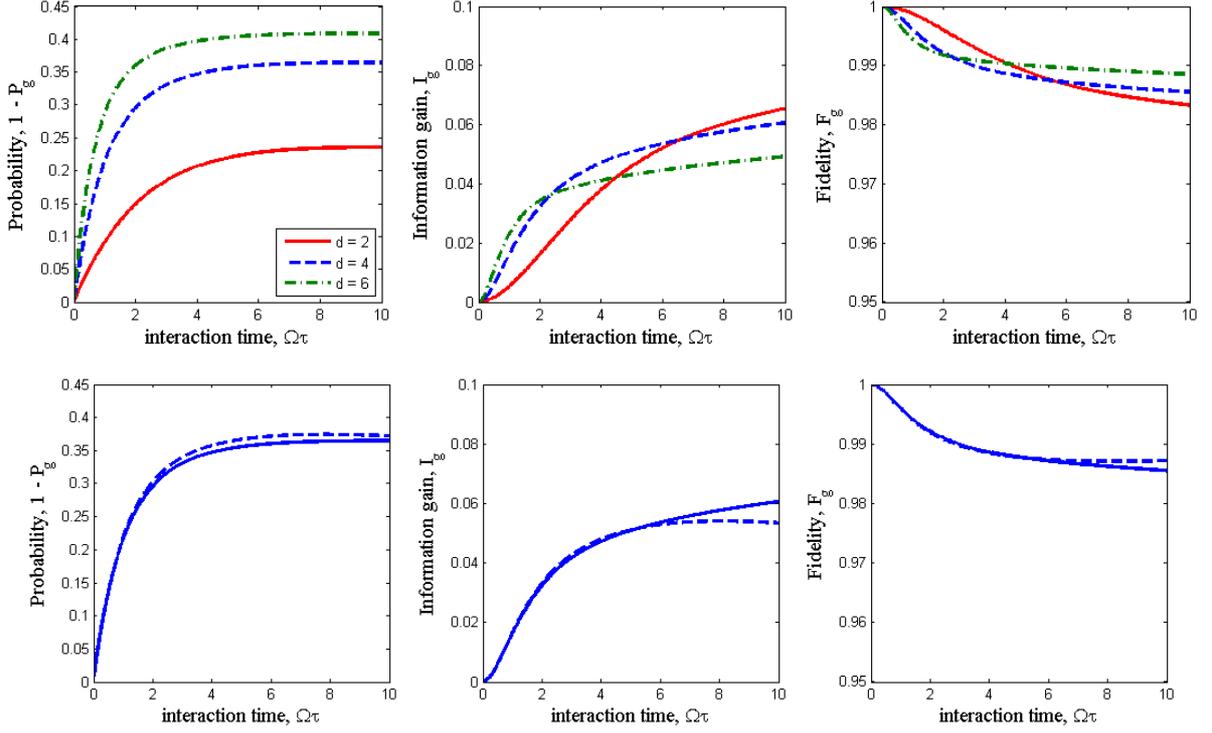

**Fig. 3** Ground state detection: probability, information gain and fidelity as a functions of interaction time: for mixed state $\rho_F = 1/d$ (top); numerical simulations (solid line) and analytical solution (dashed line) for the case $d = 4$. Weak relaxation approximation.

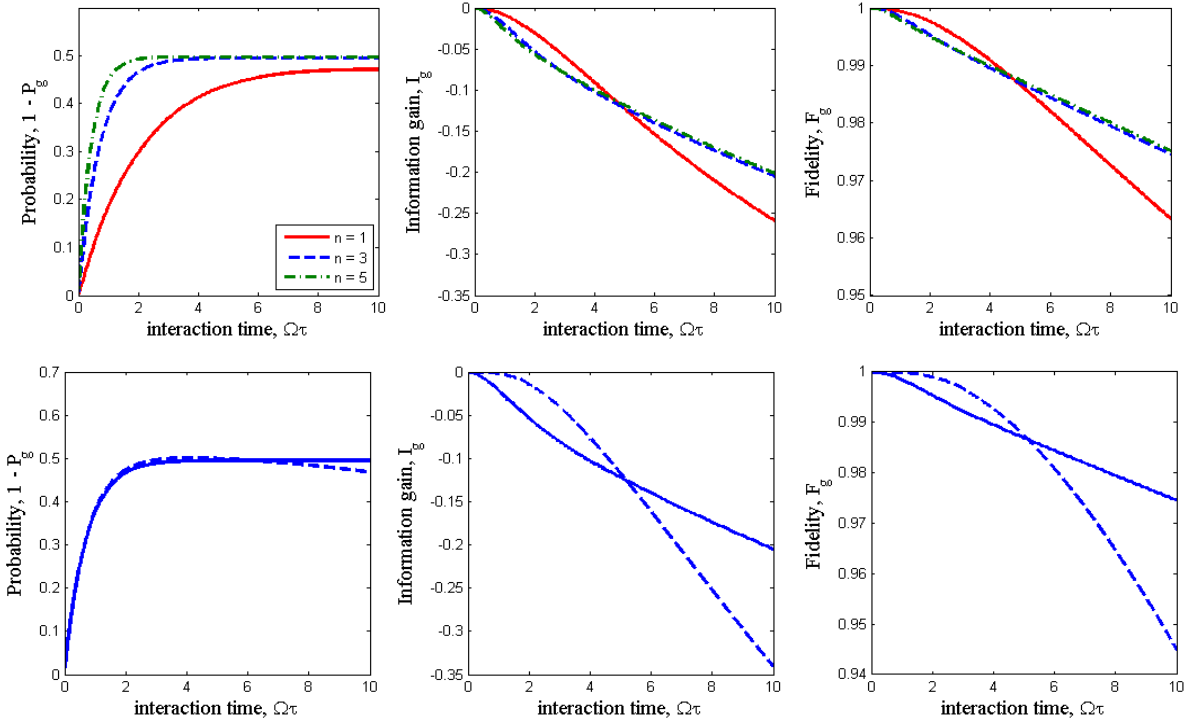

**Fig. 4** Ground state detection: probability, information gain and fidelity as functions of interaction time: for mixed state for Fock state $\rho_F = |n\rangle\langle n|$ (top); numerical simulations (solid line) and analytical solution (dashed line) for the case $n = 3$. Weak relaxation approximation.

## ACKNOWLEDGEMENTS

The work was supported by FTP "Scientific and Educational Human Resources for Innovation-Driven Russia" (contract 16.740.11.0030, grant 2012-1.2.2-12-000-1001-047), grant 11-08-00267 of RFBR and by FTP "Researches and Development in the Priority Directions Developments of a Scientific and Technological Complex of Russia 2007-2013" (contract 07.514.11.4146).